\documentclass[aps, prd, amsmath, floats, floatfix, twocolumn, nofootinbib, superscriptaddress, showpacs]{revtex4-1}
\usepackage{graphicx}
\usepackage{color}
\usepackage{units}
\usepackage{latexsym}

\newcommand{\be}{\begin{eqnarray}}
\newcommand{\ee}{\end{eqnarray}}
\newcommand{\rar}{\rightarrow}

\begin{document}

\title{Formation and evaporation of an electrically charged black hole in conformal gravity}

\author{Cosimo Bambi}
\email{bambi@fudan.edu.cn}
\affiliation{Center for Field Theory and Particle Physics and Department of Physics, Fudan University, 200433 Shanghai, China}
\affiliation{Theoretical Astrophysics, Eberhard-Karls Universit\"at T\"ubingen, 72076 T\"ubingen, Germany}

\author{Leonardo Modesto}
\email{lmodesto@sustc.edu.cn}
\affiliation{Department of Physics, Southern University of Science and Technology, Shenzhen 518055, China}

\author{Shiladitya Porey}
\email{shiladitya@g.nsu.ru}
\affiliation{Novosibirsk State University, Novosibirsk, Novosibirsk Oblast, 630090, Russia}

\author{Les\l{}aw Rachwa\l{}\,}
\email{grzerach@gmail.com}
\affiliation{Instituto de F\'{i}sica, Universidade de Bras\'{i}lia, 70910-900, Bras\'{i}lia, DF, Brazil}

\date{\today}
 
\begin{abstract}
Extending previous work on the formation and the evaporation of black holes in conformal gravity, in the present paper we study the gravitational collapse of a spherically symmetric and electrically charged thin shell of radiation. The process creates a singularity-free black hole. Assuming that in the evaporation process the charge $Q$ is constant, the final product of the evaporation is an extremal remnant with $M=Q$, which is reached in an infinite amount of time. We also discuss the issue of singularity and thermodynamics of black holes in Weyl's conformal gravity.
\end{abstract}

\maketitle


\section{Introduction}

The two most popular families of conformally invariant theories of gravity are Weyl's and Einstein's conformal gravity~\cite{cg1,cg2,v2-cg,cg3,cg4,cg5,cg6,cg7,cg8,cg9}. The former is a renormalizable theory of gravity with extra degrees of freedom. The latter contains only one physical degree of freedom, the graviton, but is not renormalizable. Both theories are invariant under a conformal transformation of the metric tensor
\be\label{eq-transf}
g_{\mu\nu} \rar \Omega^2 g_{\mu\nu} \, ,
\ee
where $\Omega = \Omega(x)$ is a function of the spacetime point. In the case of Einstein's conformal gravity, we also have a conformal compensator field $\phi$ (dilaton) and a possible action is~\cite{v2-dirac}
\be\label{eq-confEgrav}
S = - 2 \int d^4 x \, \sqrt{|g|} 
\left( \phi^2 R + 6 \, g^{\mu\nu} (\partial_\mu \phi)( \partial_\nu \phi) \right) .
\ee
This action is invariant under the conformal transformation in~(\ref{eq-transf}) if the dilaton transforms homogeneously~\cite{v2-navarro}
\be
\phi \rar \Omega^{-1} \phi \, .
\ee
If $(g_{\mu\nu}, \phi)$ is a solution of the field equations of the theory, then even $(g_{\mu\nu}^*, \phi^*)$, where
\be
g_{\mu\nu}^* = \Omega^2 g_{\mu\nu} \, , \quad \phi^* = \Omega^{-1} \phi \, ,
\ee
is a solution. Note that for $\phi = 1/\sqrt{32 \pi}={\rm const}$ we recover Einstein's gravity with the proper normalization\footnote{ In this paper we employ natural units: $c = G_N = \hbar =k_B= 1$ and a metric with signature $( - + + \, + )$.}. However, Einstein's gravity is not conformally invariant in its standard formulation. In the present paper, we will focus on Einstein's conformal gravity, but we will also show how some of our results can be extended to the Weyl gravitational theory. In particular, Sections~\ref{s-2} and \ref{s-3} are devoted to Einstein's conformal gravity, while in Section~\ref{s-new} we show that our results can be easily extended to Weyl's conformal gravity.

Einstein's gravity is invariant under general coordinate transformations. If some quantity is singular in a certain coordinate system and not in another one, then the singularity is not physical, but only an artifact of the choice of the reference frame. In this case, we speak about ``coordinate singularity'', and it is not a true singularity of the spacetime. For example, in the Schwarzschild spacetime in Schwarzschild coordinates, the metric is singular at the spherical surface $r = 2M$ describing the event horizon. However, the spacetime is perfectly regular there, and it is indeed possible to remove the singularity with a suitable change of coordinates.

Conformal gravity is invariant under both general coordinate transformations and conformal transformations. In conformal gravity, it is possible to remove the spacetime singularities present in Einstein's gravity by finding a suitable conformal transformation $\Omega=\Omega(x)$~\cite{p1,p2}. In this context, we should speak about ``conformal singularities'', because, in analogy to the coordinate singularities in Einstein's gravity, they are not true singularities of the spacetime, but only artifacts of the choice of the conformal gauge.

Note that we cannot use the same mathematical tools to study the spacetime singularities in Einstein's gravity and in conformal gravity. For example, in Einstein's gravity it is common to check the regularity of a spacetime by studying invariants like the scalar curvature and the Kretschmann scalar. However, in conformal gravity the scalar curvature and the Kretschmann scalar are not invariant any longer: they are invariant under general coordinate transformations, but not under conformal transformations, and therefore they are not associated to any physical quantity. Another example is represented by the study of time-like geodesics. In Einstein's gravity, the spacetime is geodesically incomplete if the trajectory of a massive particle cannot be extended beyond a certain point. However, in conformal gravity there are no massive particles (at least in the usual way), because they would violate the conformal symmetry of the theory.

If conformal invariance is a symmetry in Nature, it must be broken, because the Universe around us is not conformally invariant. For example, in a conformally invariant theory we cannot measure lengths and time intervals, which is definitively not our case. Conformal invariance may be spontaneously broken, and in such a case Nature has selected one of the possible vacua. Conformal gravity is an appealing theoretical framework to solve the problem of spacetime singularities when we assume that Nature can only select a physical vacuum in the class of singularity-free metrics~\cite{cg1,cg2,bobir}.

It is well known that for the theory in~(\ref{eq-confEgrav}) conformal invariance is broken at the quantum level~\cite{Fradkin:1981iu}. However, in any finite (i.e. without quantum divergences) quantum field theory of gravity, conformal invariance is preserved~\cite{p1}. This is realized in a class of weakly nonlocal gravitational theories that turn out to be finite at the quantum level and perturbatively unitary~\cite{l1,l2,l3}.

Regarding Einstein's conformal gravity, in Refs.~\cite{p1,p2} we found a family of conformal factors $\Omega$ that make the Schwarzschild solution regular everywhere. In the simplest case, which is the one considered later in this work, the conformal factor $\Omega$ reads
\be\label{eq-conf-tran}
\Omega = \left(1 + \frac{L^2}{r^2}\right)^2 \, ,
\ee
where $L$ is a parameter with the dimension of length. It is natural to assume that $L$ is either of order the Planck length $L_{\rm Pl} \sim 10^{-33}$~cm or of order the gravitational radius of the system. In the former case, observational tests may be out of reach. In the latter case, we have the constraint $L/M < 1.2$ from the X-ray spectrum of astrophysical black holes~\cite{p3,review}, and independent constraints may be obtained from future observations of gravitational waves~\cite{uz}. 

Moreover, the Schwarzschild metric rescaled by (\ref{eq-conf-tran}) is certanly a regular black hole's solution of the Weyl conformal gravity too. Indeed, any Ricci-flat spacetime in whatever Weyl invariant gravity theory turns out to be singularity-free for a suitable rescaling.

In the present paper, we continue our study on the formation and evaporation of black holes in conformal gravity extending our previous work~\cite{b1,b2,b3,shila1}. In particular, we study the gravitational collapse of a spherically symmetric and electrically charged thin shell of radiation. The outcome of the collapse is a spherically symmetric and electrically charged singularity-free black hole, which is the counterpart of the singular Reissner-Nordstr\"om solution in Einstein's gravity. We study the evaporation process and discuss our finding considering also the results of our previous work~\cite{shila1}.

\section{Gravitational collapse \label{s-2}}

Massive particles are not conformally invariant. The simplest model of gravitational collapse in conformal gravity in the symmetric phase is represented by the collapse of a thin shell of radiation. In the present paper, we want to consider the possibility of a non-vanishing electric charge.

In Einstein's gravity, the gravitational collapse of a spherically symmetric and electrically charged thin shell of radiation is described by the Vaidya-Bonnor ingoing metric~\cite{v2-vb}
\be
ds^2_{\rm V} &=& - \left(1 - \frac{2 M(v)}{r} + \frac{Q(v)}{r^2} \right) dv^2 + 2 dv dr 
\nonumber\\
&& + r^2 \left(d\theta^2 + \sin^2\theta d\phi^2\right) \, ,
\label{Vaidya_metric}
\ee
where $M(v)$ and $Q(v)$ are given, respectively, by
\be
M(v) &=& M_{0}\Theta (v-v_{0}) \, , \\
Q(v) &=& Q_{0}\Theta (v-v_{0}) \, .
\ee
Here $v$ is the ingoing null coordinate, $M_{0}$ is the ADM mass, and $\Theta$ is the unit step function (Heaviside function). The energy-momentum tensor of the thin shell of radiation is
\be\label{Tmunu}
T_{\mu\nu} &=& \frac{1}{8\pi}(R_{\mu\nu}- \frac{1}{2}R g_{\mu \nu}) \nonumber\\
&=& \left( \begin{array}{cccc}
	T_{vv} & T_{vr} & 0 & 0 \\
	T_{rv} & 0 & 0 & 0 \\
	0 & 0 & T_{\theta\theta} & 0 \\
	0 & 0 & 0 & T_{\phi\phi} \\
\end{array} \right) \, ,
\ee
where 
\be
T_{vv} &=& - \frac{Q_0 - 2 M_0 r}{8\pi r^3} \delta (v-v_{0}) \nonumber\\
&& + \frac{Q_0 \left( Q_0 - 2 M_0 r + r^2 \right)}{8\pi r^6} \Theta (v-v_{0}) \, , \nonumber\\
 T_{vr} &=& T_{rv} =  - \frac{Q_0}{8\pi r^4} \Theta (v-v_{0}) \, , \nonumber\\ 
T_{\theta\theta} &=& \frac{Q_0}{8\pi r^2} \Theta (v-v_{0}) \, , \nonumber\\
T_{\phi\phi} &=& \frac{Q_0 \sin^2\theta}{8\pi r^2} \Theta (v-v_{0}) \, .
\ee
This energy-momentum tensor follows from the metric in Eq.~\eqref{Vaidya_metric}. It is easy to check that $T_{\mu\nu}$ in~\eqref{Tmunu} is traceless, which is an important consistency requirement when it is coupled to conformal gravity.

The Vaidya-Bonnor ingoing metric~\eqref{Vaidya_metric} is a solution also in the theory~\eqref{eq-confEgrav}. However, in Einstein's gravity the metric~\eqref{Vaidya_metric} is singular: the gravitational collapse forms a central spacetime singularity, which is a singularity both in the sense of geodesic incompleteness of the  spacetime and of curvature singularity. In Einstein's conformal gravity, the metric in~\eqref{Vaidya_metric} is singular in the sense that the conformal gauge is singular, but the spacetime is regular. If we apply the conformal transformation in~\eqref{eq-conf-tran}, we obtain
\be\label{eq-vai}
ds^2 = \left(1 + \frac{L^2}{r^2}\right)^4
ds^2_{\rm V} \, .
\ee
where $ds^2_{\rm V}$ is the line element in~\eqref{Vaidya_metric}. The scalar field $\phi$ had the profile  $1/\sqrt{32 \pi}$ in Einstein's gravity, but now has the following non-constant profile,
\be
\phi = \frac{1}{\sqrt{32 \pi}} \left(1 + \frac{L^2}{r^2}\right)^{-2} \, .
\ee
The scalar invariants (with respect to general coordinate transformations) for this conformally transformed Vaidya-Bonnor metric are calculated in the appendix and are regular for $L \neq 0$ at all radii $r$ and all values of the  null ingoing coordinate $v$.

The physical picture of the gravitational collapse of the thin shell of radiation in Einstein's conformal gravity is the same as in Einstein's gravity. The shell moves along the null geodesic $v = v_0$. Inside the shell, the spacetime is conformally flat. Outside the shell, the metric is described by the conformally modified Reissner-Nordstr\"om metric
\be\label{eq-rn}
ds^2 = \left(1 + \frac{L^2}{r^2}\right)^4 ds^2_{\rm RN} \, ,
\ee
where $ds^2_{\rm RN}$ is the line element of the standard Reissner-Nordstr\"om metric
\be
ds^2_{\rm RN}&=&-\left(1-\frac{2M}{r}+\frac{Q^2}{r^2}\right)dt^2 \nonumber\\
&& +\left(1-\frac{2M}{r}+\frac{Q^2}{r^2}\right)^{-1}dr^2
+r^2d\Omega^2\,.
\ee
The event horizon forms when the shell crosses the radius $r_{\rm H} = M + \sqrt{M^2 - Q^2}$, as in Einstein's gravity. However, now no singularity is formed at the center (even in the sense of curvature invariants under general coordinate transformations). The energy-momentum tensors of the radiation~\eqref{Tmunu} and of the scalar field $\phi$ diverge at $r \to 0$, as well as their sum, but these are not invariant quantities as they depend on the choice of the coordinate system.

\section{Black hole evaporation \label{s-3}}

The thermodynamical properties of a black hole depend on the metric on and outside its event horizon. We can thus study the evaporation process of the newly born black hole with the metric in~\eqref{eq-rn} instead of the metric in~\eqref{eq-vai}, which significantly simplifies our calculations.

The line element of a static and spherically symmetric black hole spacetime can always be written as
\be
ds^2 = g_{tt} dt^2 + g_{rr} dr^2 + r^2 \left(d\theta^2 + \sin^2\theta d\phi^2\right) \, , 
\label{metricgen}
\ee
where the metric functions $g_{tt}$ and $g_{rr}$ are independent of the coordinates $t$, $\theta$, and $\phi$. With such a choice of the metric tensor, the surface gravity is~\cite{visser}
\be
\kappa =- \lim_{r \rar r_{\rm H}}  \frac{1}{2 \sqrt{|g_{tt} g_{rr}|}} \frac{\partial g_{tt}}{\partial r} \, ,
\label{kappagen}
\ee
where $r_{\rm H}$ is the radial coordinate of the event horizon. From the surface gravity, we can evaluate the Hawking temperature of the black hole
\be
T_{\rm H} = \frac{\kappa}{2 \pi} \, .
\ee

With the modified Reissner-Nordstr\"om metric in~\eqref{eq-rn}, the Hawking temperature of the black hole is
\be
\hspace{-0.8cm}
T_{\rm H} &=&\frac{L^2 \left(r_{\rm H} (9 M - 4 r_{\rm H}) - 5 Q^2\right)
+ r_{\rm H}^2 \left(M r_{\rm H} - Q^2\right)}{2 \pi  r_{\rm H}^3 
\left(L^2 + r_{\rm H}^2\right)} \nonumber\\
&=& \frac{M r_{\rm H} - Q^2}{2 \pi  r_{\rm H}^3} \, ,
\ee
Note that $T_{\rm H}$ is independent of $L$, and thus identical to the Hawking temperature of the standard Reissner-Nordstr\"om black hole. Such a result had to be expected
because the surface gravity at the event horizon is conformally invariant. The surface area at the event horizon, $A_{\rm H}$, is instead different if $L \neq 0$
\be
A_{\rm H} &=& \lim_{r \rar r_{\rm H}} \int_0^\pi \int_0^{2\pi} 
\sqrt{g_{\theta\theta} g_{\phi\phi}} \, d\theta \, d\phi \nonumber\\
&=& \frac{4 \pi  \left(L^2 + r_{\rm H}^2\right)^4}{r_{\rm H}^6} \, .
\ee
Even this result could be expected, as in the symmetric phase we cannot perform any measurement of length (and therefore area), so $A_{\rm H}$ is not a physical quantity if the theory is explicitly conformally invariant. This, in turn, means that varying the value of $L$ we change the value of $A_{\rm H}$ too.

Since the Hawking temperature $T_{\rm H}$ is invariant under conformal transformations, the black hole entropy is  invariant as well. Integrating the expression $dS/dM = 1/T_{\rm H}$ and neglecting the constant $-2\pi Q^2$, we find
\be
S(M) = 2\pi M r_{\rm H} \, .
\label{entropy}
\ee
However, the black hole luminosity (or mass-loss rate) is different because the surface area $A_{\rm H}$ is not invariant under conformal transformations. Again, this had to be expected, as the measurement of the luminosity requires the measurement of time intervals, which are not invariant under conformal transformations.

\subsection{Canonical ensemble}

The canonical statistical ensemble is defined by the condition that the energy (mass) is constant in the statistical equilibrium state. The black hole luminosity is
\be\label{eq-dmdt}
L_{\rm H} = - \frac{dM}{dt} = \sigma A_{\rm H} T^4_{\rm H} \, ,
\label{luminosity}
\ee
where $\sigma$ is a Stefan-Boltzmann-like constant that depends on the particle content of the theory and on the black hole mass~\cite{page1,page2}. Plugging the expressions of $A_{\rm H}$ and $T_{\rm H}$, we find
\be
L_{\rm H} = \frac{\sigma  \left(L^2 + r_{\rm H}^2\right)^4 
\left(M r_{\rm H} - Q^2\right)^4}{4 \pi ^3 r_{\rm H}^{18}} \, .
\label{microLconst}
\ee

Let us assume that the electric charge $Q$ is constant during the evaporation process. For example, this is the case if the black hole can only emit uncharged particles. In such a case, the final product of the evaporation process is an extremal remnant with $M=Q$.

If we write $L_{\rm H}$ as $dM/dt$ in Eq.~(\ref{luminosity}), and we integrate, we find the evaporation time
\be\label{eq-ev-time3}
t_{\rm evap} = \int_Q^{M_0}  \frac{dM}{L_{\rm H}} \,.
\ee
For $L=0$, we recover the standard result for a Reissner-Nordstr\"om black hole in the canonical ensemble
\be
\bar t_{\rm evap} = \int_Q^{M_0} 
\frac{4\pi^3 r_{\rm H}^{10} dM}{\sigma \left(M r_{\rm H}-Q^2 \right)^4 } \, ,
\ee
If we assume $L = {\rm const}$, we find
\be
\bar t_{\rm evap} = \int_Q^{M_0} 
\frac{4\pi^3 r_{\rm H}^{18} dM}{\sigma \left( L^2+r_{\rm H}^2\right)^4 
\left(M r_{\rm H}-Q^2 \right)^4} \, .
\ee
If $L$ is proportional to the black hole mass, say $L=\alpha M$ where $\alpha$ is some constant, we have
\be
\bar t_{\rm evap} = \int_Q^{M_0} \frac{4\pi^3 r_{\rm H}^{18} dM}{\sigma \left( \alpha^2 M^2+r_{\rm H}^2\right)^4 \left(M r_{\rm H}-Q^2 \right)^4} \, .
\ee 
In the uncharged case, studied in \cite{shila1}, the relation of proportionality $L=\alpha M$ was motivated by the fact that it solved the integral equations of evaporation. However, in the present charged case, we take a possible phenomenological assumption that the length scale $L$ should depend on the characteristic length scales present in the original (non-rescaled) solution of the theory. Here we choose it to be directly proportional to the mass of the black hole $M$. Another possible choice would be to assume that it is proportional to the charge $Q$ or to some combination of the two.

We see that all evaporation times are infinite due to the behaviour near the remnant. In the standard Reissner-Nordstr\"om case, the expansion of the integrand in Eq.~(\ref{eq-ev-time3}) is
\be
\frac{\pi ^3 Q^4}{\sigma  (M-Q)^2}+O(M-Q)^{3/2}\,.
\ee
In the conformally modified case with $L={\rm const}\neq 0$, we have
\be
\frac{\pi ^3 Q^{12}}{\sigma  \left(L^2+Q^2\right)^4 (M-Q)^2}+O(M-Q)^{3/2} \, .
\ee
When $L=\alpha M$, we have 
\be
\frac{\pi ^3 Q^4}{\left(\alpha ^2+1\right)^4 \sigma  (M-Q)^2}+O(M-Q)^{3/2}\,.
\ee
We thus see that all these evaporation times diverge and a massive extremal remnant with $M=Q$ remains.

\vspace{0.5cm}

\subsection{Micro-canonical ensemble}

In the micro-canonical ensemble the statistical system is kept in contact with a thermal reservoir, so the temperature $T$ is fixed~\cite{casadio1,casadio2, hossenfelder}. The evaporation rate in micro-canonical ensemble is~\cite{hossenfelder}
\be
\frac{dM}{dt}=-{\cal E}A\,,
\ee
where ${\cal E}$ is the energy density of the radiation field surrounding the evaporating black hole. This reads in full generality
\be
{\cal E}=\frac{\sigma}{3}e^{-S(M)}\int_{Q}^{M}e^{S(x)}(M-x)^{3}dx \, ,
\ee
and in our case it becomes
\be
{\cal E} = \frac{\sigma}{3} e^{-2 \pi M r_{\rm H}} \int_Q^M e^{2 \pi x r_{\rm H} (x)} (M-x)^3 dx \, ,
\ee
where $r_{\rm H} (x) = x + (x^2 - Q^2)^{1/2}$. The mass-loss rate is
\be
\frac{dM}{dt} &=& - \frac{4\pi\left(L^2+r_{\rm H}^2\right)^4}{r_{\rm H}^6} 
\frac{\sigma}{3} e^{-2 \pi  M r_{\rm H}} 
\nonumber\\ && \,\,\,\,\,
\int_Q^M e^{2 \pi x r_{\rm H}(x)} (M-x)^3 dx \, .
\ee

In the micro-canonical ensemble, the expressions for the evaporation time are more complicated because they contain a double integral. In Einstein's gravity with $L=~\!\!0$~we have
\begin{widetext}
\be
\bar t_{\rm evap} = \int_Q^{M_0} \frac{dt}{dM} dM = \int_Q^{M_0} \frac{dM}{4\pi r_{\rm H}^8} \frac{3}{\sigma} e^{2 \pi M r_{\rm H}} \frac{1}{\int_Q^M e^{2 \pi x r_{\rm H}(x)} (M-x)^3 dx} \, .
\ee
In the conformally modified case with $L=\rm const$ we have
\be
\bar t_{\rm evap} = \int_Q^{M_0}\frac{r_{\rm H}^6 dM}{4\pi\left(L^2 +r_{\rm H}^2\right)^4} \frac{3}{\sigma} e^{2 \pi M r_{\rm H}} \frac{1}{\int_Q^M e^{2 \pi x r_{\rm H} (x)} (M-x)^3 dx} \, . 
\ee
If $L=\alpha M$, we find
\be
\bar t_{\rm evap} = \int_Q^{M_0} \frac{r_{\rm H}^6 dM}{4\pi\left(\alpha^2 M^2+r_{\rm H}^2\right)^4} \frac{3}{\sigma} e^{2 \pi M r_{\rm H}} \frac{1}{\int_Q^M e^{2 \pi x r_{\rm H}(x)} (M-x)^3 dx} \, . 
\ee
\end{widetext}

The evaporation time is also divergent in the micro-canonical ensemble. Here the conformal rescaling does not help because this divergence comes entirely from the divergence of the energy density $\cal E$ near the extremality $M=Q$. Expanding the integral, we find that $\cal E$ scales like
\be
 {\cal E} = \frac{1}{12(M-Q)^4} +O(M-Q)^{-3}
 \ee
and is independent of $L$. The mass-loss rate is vanishing exactly at the extremal point in both canonical and micro-canonical ensembles. The area surface of the horizon is regular in this limit, while it is important for the canonical ensemble that the Hawking temperature scales like
 \be
 \frac{\sqrt{M-Q}}{\sqrt{2} \pi  Q^{3/2}}+O(M-Q)\,.
 \ee
 This implies that the mass-loss rate near the extremal point vanishes and its expansion is
 \be
  \frac{(M-Q)^2}{4 \pi ^4 Q^6}+O(M-Q)^3
  \ee
again independently of the conformal rescaling.

More generally, if we have (or not) a remnant as the final state of the evaporation process and the expression for the mass-loss rate is an analytic function of the difference between the actual mass of the black hole and the mass of the remnant (or zero if there is no remnant to be left), then the evaporation time will always be infinite. If $dM/dt$ was a function whose behaviour is with the power exponent on the difference smaller than 1 (so non-analytic), then the evaporation time would be finite.

We notice that in the conformal gravity considered in this section as well as in Einstein's gravity the evaporation times of charged black holes not possessing naked singularity ($M>Q$) diverge in all ensembles here studied. This is an evidence for the analogue of the third law of thermodynamics for black holes since the extremal black holes ($M=Q$) have vanishing Hawking temperature $T_{\rm H}=0$. To reach them by any thermodynamical process (like Hawking evaporation) requires infinite amount of time so they are unattainable thermodynamical states. This situation is changed in Weyl's conformal gravity as we discuss below.

\section{Black holes in Weyl's conformal gravity} \label{s-new}

It is also interesting to study the situation of electrically (or magnetically) charged black hole's solutions in Weyl's conformal gravity that is defined by the unique local conformally invariant action in a four-dimensional spacetime without need of a dilaton field
\be
S_{C^2}=-\alpha\!\int\!d^4x\sqrt{|g|}C_{\mu\nu\rho\sigma}C^{\mu\nu\rho\sigma} \, , 
\ee
where $C_{\mu\nu\rho\sigma}$ is the Weyl tensor (of conformal curvature). In our case the field equations tell that the Bach tensor must be equal to the energy-momentum tensor of the electromagnetic field surrounding a static charged black hole, namely
\be
B^{\mu\nu}=\frac{\delta S_{C^2}}{\delta g_{\mu\nu}}=- 2 \sqrt{|g|}T^{\mu\nu}_{\rm el}\,.
\label{Mansatz}
\ee

This is now a four-derivative theory and due to the non-linearities of gravitation it is difficult to find exact solutions in our setup. However, already in 1990 Mannheim and Kazanas found a new class of solutions \cite{Mannheim:1988dj,Mannheim:1990ya,MannheimAnnals} pertaining to our problem. In what follows we discuss their static spherically symmetric solutions. In particular, we will properly rescale these metrics in order to end up with regular spacetimes and, afterwords, we will study the black hole evaporation in this setup.

 \subsection{Singularity-free black hole}

In conformal gravity, multiple solutions are physically equivalent if they differ by an overall conformal factor. Therefore, we can assume the latter to be $\Omega = 1$ and, following \cite{Mannheim:1988dj,Mannheim:1990ya,MannheimAnnals}, we can look for solutions with ansatz:
\be
ds^2=-B(r)dt^2+B(r)^{-1}dr^2+r^2d\Omega^2 \, .
\ee
The EOM can be integrated exactly and the function $B(r)$ is:
\be
B(r)=w+\frac{u}{r}+ \tilde{v} r-kr^2 \, ,
\label{Msol}
\ee
where $w,u, \tilde{v},k$ are integration constants and the following relation between them is satisfied
\be
w^2-1-3 u \tilde{v}=\frac{3Q^2}{8\alpha} \, .
\ee

For the sake of simplicity, we further choose the case $\tilde{v}=0$ and, in order to have solutions asymptotically flat, we also take the parameter $k$ to vanish. This is an exact solution for a static black hole with the charge $Q$. The only drawback of this solution for the metric, understood in standard (non-conformal) general relativity, is that it possesses a singularity at $r=0$, similarly to the case of the standard Reissner-Nordstr\"om spacetime. This is both curvature and geodesic singularity. However, in a theory explicitly enjoying conformal symmetry we can easily remove this singularity by exploiting the freedom given by the overall conformal factor in the metric. For example with the conformal factor
\be
\Omega^2(r)=1 + \frac{L^4}{r^4}
\label{cf1}
\ee
we get a complete resolution of singularity.

In conformal gravity there is no singularity in $r=0$ because one can check that standard GR-invariants (like the Kretschmann scalar) are regular. Moreover, for any probe (both massless and massive: conformally or minimally coupled to gravity) the amount of an affine parameter on their geodesics needed to reach the point $r=0$ is infinite. Hence, the point $r=0$ is unreachable. This resolution of the singularities proceeds exactly the same way in any theory of conformal gravity \cite{p2}.

\subsection{Black hole evaporation}

Since the conformal factor \eqref{cf1}, which makes the solution completely regular, is a bit complicated, we would like to present here thermodynamical analysis for the simplified solution \eqref{Msol}  when $\tilde{v}=0$. (This case is the most similar, but still different from RN solution.) In this particular case to determine the position of the horizon we need to solve a condition $B(r)=0$ for values of the radial coordinate $r$. With $\tilde{v}=0$ we solve easily the linear equation and we find $r_{\rm H}=-\nicefrac{u}{w}$. We assume that the radius $r_{\rm H}$ takes a positive value. Here we note that the analysis for the case with non-vanishing $v$ and $k$ is hampered by the fact that we would have to solve quadratic or cubic equation respectively to determine the position of the horizon. Another important remark is that to understand the causal structure of the spacetime we need only to study the trajectories of light rays so only the conformal structure of spacetime matters.

Although the analysis can be performed in full generality using the formulas \eqref{metricgen}, \eqref{kappagen} and further, here we concentrate on the case of the metric \eqref{Mansatz}
with
\be
B(r)=\sqrt{\frac{3Q^2}{8\alpha}+1}+\frac{u}{r}\,,
\ee
 where $u$ is an arbitrary negative parameter. The precise location of the spherical surface of the horizon is at the radius
 \be
 r_{\rm H}=-u\sqrt{\frac{8\alpha}{3Q^2+8\alpha}}\,.
 \ee
 We now very briefly report the results of the analysis, whose main steps were laid down above and also in the previous paper \cite{shila1}. The Hawking temperature reads here
 \be
 T_{\rm H}=-\frac{3 Q^2+8 \alpha }{32 \pi  u \alpha }\,,
 \ee
 while the entropy is (up to an additive constant)
  \be
S(M)= \frac{32 \pi  M^2 \alpha }{3 Q^2+8 \alpha }\,,
 \ee
 where we also recalled the relation between the mass $M$ and the parameter $u$, namely $u=-2M$ (valid for $v=0$).
 The luminosity in the canonical statistical ensemble is computed as
   \be
L_{\rm H}=\frac{\sigma  \left(3 Q^2+8 \alpha \right)^3}{2^{17} \pi ^3 M^2 \alpha
   ^3}\,,
 \ee
 so the time of evaporation is finite and given by
 \be
 t_{\rm evap}=\frac{2^{17} \pi ^3 M^3 \alpha ^3}{3 \sigma  \left(3 Q^2+8
   \alpha \right)^3}\,.
 \ee
Here the evaporation process does not leave any remnant provided that we do not start with the extremal black hole for which $3Q^2=-8\alpha$ (existing only for $\alpha<0$). The reason for this behaviour is that here the condition for extremality does not depend on the mass of the black hole, but only on the constant charge $Q$ and the parameter of the theory $\alpha$.

\begin{widetext}
In the micro-canonical ensemble we have the following expression for the luminosity
 \be
 L_{\rm H}=-\frac{dM}{dt}&=&-\frac{\sigma M^2  e^{-\frac{32 \pi  \alpha M^2}{3 Q^2+8 \alpha }}}{48 \pi  \alpha  \left(3 Q^2+8 \alpha \right)} \left\{2\pi \sqrt{2\alpha}  \sqrt{3 Q^2+8 \alpha } M
   \left[8 \left(3-8 \pi  M^2\right) \alpha +9 Q^2\right]
   \text{erfi}\left(\frac{4 \sqrt{2 \pi \alpha } M }{\sqrt{3
   Q^2+8 \alpha }}\right)\right.+\nonumber\\
&&   \left.\left(3 Q^2+8 \alpha \right)
   \left(e^{\frac{32 \pi \alpha M^2  }{3 Q^2+8 \alpha }} \left[8
   \left(4 \pi  M^2  -1\right)\alpha-3 Q^2\right]-96 \pi  M^2
   \alpha +3 Q^2+8 \alpha \right)\right\}\,.
 \ee
 \end{widetext}
 With this at hand we could compute the evaporation time in this ensemble. However, it is found to be divergent exactly near the point $M=0$. This is a consequence of the fact how the energy density $\cal E$ behaves for almost massless black holes. We find very similarly to the cases studied in \cite{shila1} that the behaviour of the mass-loss rate is
\be
\frac{dM}{dt}=-\frac{32 \pi  \alpha  \sigma }{3 \left(3 Q^2+8 \alpha \right)}M^6+O\left({M^5}\right)\,.
\ee
Hence the evaporation time is here formally infinite. The same dependence proportional to $M^6$ was found in the dynamics of Schwarzschild black hole showing its universality across different models of black hole evaporation.

\section{Concluding remarks \label{s-5}}

In this paper, we analyzed the formation and evaporation process of a spherically symmetric and electrically charged black hole in conformal gravity. The black hole is created by the collapse of a spherically symmetric and electrically charged thin shell of radiation. We worked with a singularity-free spacetime in the spontaneously broken phase of conformal invariance. Curvature invariants like the scalar curvature, the square of the Ricci tensor, and the Kretschmann scalar remain finite over the whole spacetime. The center of the black hole at $r=0$ can be reached neither by massive particles in a finite proper time, nor by massless particles with a finite value of their affine parameter.

We studied the evaporation process in the canonical and micro-canonical ensembles in details in Einstein's conformal gravity, assuming the possibility of emission of only uncharged particles. In both ensembles, the evaporation process always produces an extremal remnant with $M=Q$, which is approached in an infinite amount of time. In Weyl's conformal gravity we studied thermodynamics of Mannheim's exact black hole solutions. We found that the evaporation time is there finite in the canonical and infinite in the micro-canonical ensemble, qualitatively very similarly to the case of Schwarzschild black hole in Einstein's gravity.

It is worth noting that conformal invariance can fix the singularity issue even in the presence of the Maxwell field in its own way. While the electric potential remains singular at the center $r=0$, this is not a problem because no massive (massless) particle can reach the center in a finite proper time (finite value of their affine parameter). This point is thus unattainable and the singularity of the electric potential has no physical implications.


\begin{acknowledgments}
C.B. acknowledges support from the National Natural Science Foundation of China (Grant No.~U1531117), Fudan University (Grant No.~IDH1512060), and the Alexander von Humboldt Foundation. S.P. thanks the Department of Physics at Fudan University for hospitality during his visit.
\end{acknowledgments}


\newpage

\begin{widetext}

\section*{Appendix: Scalar invariants of Vaidya spacetime \label{s-a}}

Scalar invariants in the conformally modified Vaidya spacetime are everywhere regular and do not diverge at $r = 0$ for $L \neq 0$. To prove this assertion, we list some scalar invariant functions (containing up to four derivatives of the metric). The Kretschmann scalar is

\be
R_{\mu\nu\rho\sigma} R^{\mu\nu\rho\sigma} &=& \frac{8 r^8 \left(-6 M_0 r \theta \left(v-v_0\right) \left(2 \left(3 L^8 \left(35 Q_0^2+24 r^2\right)+8 L^6 \left(7 Q_0^2 r^2-2 r^4\right)+2 L^4 \left(59 Q_0^2
	r^4+36 r^6\right)     \right) \right) \right)    }{\left(L^2+r^2\right)^{12}}                \nonumber\\
&+&    \frac{8 r^8 \left(-6 M_0 r \theta \left(v-v_0\right) \left(2 \left(      8 L^2 Q_0^2 r^6+Q_0^2 r^8\right)-M_0 \left(177 L^8 r+36 L^6 r^3+182 L^4 r^5+4 L^2 r^7+r^9\right)\right)       \right)}{\left(L^2+r^2\right)^{12}}        \nonumber\\
&+&        \frac{8 r^8 \left(      L^8 \left(387 Q_0^4+492 Q_0^2 r^2+184
	r^4\right)+4 L^6 \left(84 Q_0^4 r^2+11 Q_0^2 r^4-28 r^6\right)+2 L^4 \left(239 Q^4_0 r^4+266 Q^2_0 r^6+92 r^8\right)       \right)}{\left(L^2+r^2\right)^{12}}          \nonumber\\
&+&    \frac{8 r^8 \left(         4 L^2 \left(14 Q^4_0 r^6+5 Q^2_0 r^8\right)+7
	Q^4_0 r^8\right)}{\left(L^2+r^2\right)^{12}}
\, .
\ee
The square of the Ricci tensor is
\be
R_{\mu\nu} R^{\mu\nu} &=& \frac{4 r^8 \left(-48 L^2 M_0 r \theta \left(v-v_0\right) \left(6 L^6 \left(11 Q^2_0+7 r^2\right)+L^4 \left(57 Q_0^2 r^2+16 r^4\right)+L^2 \left(40 Q^2_0 r^4+22
	r^6\right)   \right) \right)    }{\left(L^2+r^2\right)^{12}}       \nonumber\\
&-& \frac{4 r^8 \left(-48 L^2 M_0 r \theta \left(v-v_0\right) \left(     6 M_0 \left(9 L^6 r+6 L^4 r^3+5 L^2 r^5\right)+Q^2_0 r^6\right)+L^8 \left(981 Q^4_0+1212 Q^2_0 r^2+400 r^4\right) \right)}{\left(L^2+r^2\right)^{12}}    \nonumber\\
&+&  \frac{4 r^8 \left(      4 L^6 \left(258 Q_0^4 r^2+179 Q^2_0
	r^4+8 r^6\right)+L^4 \left(658 Q_0^4 r^4+676 Q_0^2 r^6+208 r^8\right)+4 L^2 \left(8 Q_0^4 r^6+5 Q_0^2 r^8\right)+Q_0^4 r^8\right)}{\left(L^2+r^2\right)^{12}}  \nonumber\, ,
\ee
while the scalar curvature is
\be
R &=& -\frac{24 L^2 r^4 \left(-4 M_0 r \theta \left(v-v_0\right) \left(2 L^2+r^2\right)+L^2 \left(5 Q_0^2+3 r^2\right)+3 Q_0^2
	r^2+r^4\right)}{\left(L^2+r^2\right)^6}       \, .
\ee

The modified Vaidya metric has thus no curvature singularities. As in the case of the modified Schwarzschild metric~\cite{p1}, one can also check that this spacetime is \emph{not} geodesically incomplete at $r=0$ because the origin $r=0$ is reached with an infinite value of the geodesic affine parameter. For example massive particles never reach (or come out from) the point $r=0$ in a finite amount of their proper time.
\end{widetext}


\begin{thebibliography}{99}
  
\bibitem{cg1} 
  F.~Englert, C.~Truffin and R.~Gastmans,
  Nucl.\ Phys.\ B {\bf 117}, 407 (1976).
  
\bibitem{cg2} 
  J.~V.~Narlikar and A.~K.~Kembhavi,
  Lett.\ Nuovo Cim.\  {\bf 19}, 517 (1977).    
  
\bibitem{v2-cg} 
  R.~J.~Riegert,
  Phys.\ Rev.\ Lett.\  {\bf 53}, 315 (1984).
  
\bibitem{cg3} 
  G.~'t Hooft,
  Found.\ Phys.\  {\bf 41}, 1829 (2011)
  [arXiv:1104.4543 [gr-qc]].  
  
\bibitem{cg4} 
  G.~'t Hooft,
  Subnucl.\ Ser.\  {\bf 47}, 251 (2011)
  [arXiv:0909.3426 [gr-qc]].

\bibitem{cg5} 
  P.~D.~Mannheim,
  Found.\ Phys.\  {\bf 42}, 388 (2012)
  [arXiv:1101.2186 [hep-th]].

\bibitem{cg6} 
  I.~Bars, P.~Steinhardt and N.~Turok,
  Phys.\ Rev.\ D {\bf 89}, 043515 (2014)
  [arXiv:1307.1848 [hep-th]].

\bibitem{cg7} 
  I.~Bars, S.~H.~Chen and N.~Turok,
  Phys.\ Rev.\ D {\bf 84}, 083513 (2011)
  [arXiv:1105.3606 [hep-th]].

\bibitem{cg8} 
  I.~J.~Araya, I.~Bars and A.~James,
  arXiv:1510.03396 [hep-th].

\bibitem{cg9} 
  P.~Dominis Prester,
  arXiv:1309.1188 [hep-th].    
  
\bibitem{v2-dirac} 
  P.~A.~M.~Dirac,
  Proc.\ Roy.\ Soc.\ Lond.\ A {\bf 333}, 403 (1973).
  
\bibitem{v2-navarro} 
  I.~Navarro and K.~Van Acoleyen,
  JHEP {\bf 0508}, 019 (2005)
  [hep-th/0504086].
  
\bibitem{p1} 
  L.~Modesto and L.~Rachwal,
  arXiv:1605.04173 [hep-th].  
  
\bibitem{p2} 
  C.~Bambi, L.~Modesto and L.~Rachwal,
  JCAP {\bf 1705}, 003 (2017)
  [arXiv:1611.00865 [gr-qc]].
  
\bibitem{bobir} 
  B.~Toshmatov, C.~Bambi, B.~Ahmedov, A.~Abdujabbarov and Z.~Stuchlik,
  Eur.\ Phys.\ J.\ C {\bf 77}, 542 (2017)
  [arXiv:1702.06855 [gr-qc]].    
  
\bibitem{Fradkin:1981iu} 
  E.~S.~Fradkin and A.~A.~Tseytlin,
  Nucl.\ Phys.\ B {\bf 201}, 469 (1982).  
  
\bibitem{l1} 
L. Modesto and L. Rachwal, JHEP {\bf 1512}, 173 (2015) [arXiv:1506.08619 [hep-th]].
  
\bibitem{l2} 
L. Modesto and L. Rachwal, Nucl.\ Phys.\ B {\bf 889}, 228 (2014) [arXiv:1407.8036 [hep-th]]; 

\bibitem{l3} 
L. Modesto and L. Rachwal, Nucl.\ Phys.\ B {\bf 900}, 147 (2015) [arXiv:1503.00261 [hep-th]].  
  
\bibitem{p3} 
  C.~Bambi, Z.~Cao and L.~Modesto,
  Phys.\ Rev.\ D {\bf 95}, 064006 (2017)
  [arXiv:1701.00226 [gr-qc]].  
  
\bibitem{review} 
  C.~Bambi,
  Rev.\ Mod.\ Phys.\  {\bf 89}, 025001 (2017)
  [arXiv:1509.03884 [gr-qc]].  
  
\bibitem{uz} 
  B.~Toshmatov, C.~Bambi, B.~Ahmedov, Z.~Stuchlik and J.~Schee,
  Phys.\ Rev.\ D {\bf 96}, 064028 (2017)
  [arXiv:1705.03654 [gr-qc]].  
  
\bibitem{b1} 
  C.~Bambi, D.~Malafarina and L.~Modesto,
  Phys.\ Rev.\ D {\bf 88}, 044009 (2013)
  [arXiv:1305.4790 [gr-qc]].  
  
\bibitem{b2} 
  C.~Bambi, D.~Malafarina and L.~Modesto,
  Eur.\ Phys.\ J.\ C {\bf 74}, 2767 (2014)
  [arXiv:1306.1668 [gr-qc]].  
  
\bibitem{b3} 
  C.~Bambi, D.~Malafarina and L.~Modesto,
  JHEP {\bf 1604}, 147 (2016)
  [arXiv:1603.09592 [gr-qc]].    

\bibitem{shila1} 
  C.~Bambi, L.~Modesto, S.~Porey and L.~Rachwal,
  JCAP {\bf 1709}, no. 09, 033 (2017)
  [arXiv:1611.05582 [gr-qc]].  
  
\bibitem{v2-vb} 
  W.~B.~Bonnor and P.~C.~Vaidya,
  Gen.\ Rel.\ Grav.\  {\bf 1}, 127 (1970).

\bibitem{visser} 
  M.~Visser,
  Phys.\ Rev.\ D {\bf 46}, 2445 (1992)
  [hep-th/9203057].  
  
\bibitem{page1} 
  D.~N.~Page,
  Phys.\ Rev.\ D {\bf 13}, 198 (1976).  
  
\bibitem{page2} 
  D.~N.~Page,
  Phys.\ Rev.\ D {\bf 14}, 3260 (1976).    

\bibitem{casadio1} 
  R.~Casadio and B.~Harms,
  Phys.\ Lett.\ B {\bf 487}, 209 (2000)
  [hep-th/0004004].  

\bibitem{casadio2} 
  R.~Casadio and B.~Harms,
  Phys.\ Rev.\ D {\bf 64}, 024016 (2001)
  [hep-th/0101154].
  
\bibitem{hossenfelder} 
  S.~Hossenfelder,
  hep-ph/0412265.  
  

\bibitem{Mannheim:1988dj} 
  P.~D.~Mannheim and D.~Kazanas,
  Astrophys.\ J.\  {\bf 342}, 635 (1989).
  
\bibitem{Mannheim:1990ya} 
  P.~D.~Mannheim and D.~Kazanas,
  Phys.\ Rev.\ D {\bf 44}, 417 (1991).
  
  
\bibitem{MannheimAnnals}
P.~D.~Mannheim, Annals N.\ Y.\ Acad.\ Sci.\  {\bf 631}, no. 1, 194 (1991).





\end{thebibliography}
\end{document}